\documentclass[a4paper,12pt]{article}
\usepackage{amsfonts}
\usepackage{amsmath}

\input{tcilatex}

\begin{document}

\title{Resolvent Operator Transformations and Bound--State Solutions for
Confluent Natanzon Potentials}
\author{S.--A. Yahiaoui\thanks{%
E-mail: s\_yahiaoui@mail.univ-blida.dz}, M. Bentaiba\thanks{%
corresponding author E-mail: bentaiba@mail.univ-blida.dz} \\
LPTHIRM, D\'{e}partement de Physique, Facult\'{e} des Sciences,\\
Universit\'{e} Saad DAHLAB de Blida, Blida, Algeria}
\maketitle

\begin{abstract}
An algebraic method of constructing the confluent Natanzon potentials
endowed with position--dependent mass is presented. This is possible by
identifying the \textit{scaling} resolvent operator (Green's function) to
nonrelativistic position--dependent mass Schr\"{o}dinger equation.

\textbf{PACS: }03.65.Fd, 02.20.-a

\textbf{Keyword}: Resolvent operator; position--dependent mass.
\end{abstract}

\section{Introduction}

The (confluent) Natanzon potentials [1,2] are known nowadays as a general
class of solvable potentials related to the (confluent) hypergeometric
functions and cover all well--known potentials in quantum mechanics for
which an exact solutions can be found [3]. The (Radial) harmonic oscillator
potential [4], the Coulomb potential [4] and the Morse potential [5] may all
be considered as particular cases of the confluent hypergeometric
potentials, while the hypergeometric potentials consist of the P\"{o}%
schl--Teller potential [6], Eckart potential [7], Rosen--Morse potential
[8], etc.

However, different series of papers have been devoted in recent years to the
study of the dynamical symmetries [9--17]. It was proved that the bound and
scattering states of these potentials are related to the representation of
the $\mathfrak{so(}2,1\mathfrak{)}$ $\left[ \sim \mathfrak{su}\left(
1,1\right) \sim \mathfrak{sl}\left( 2,%
\mathbb{R}
\right) \right] $ Lie algebras; these algebras play the role of dynamical
algebras. They are used as a \textit{spectrum generating algebra} [9--13]
and as a \textit{potential algebra} [14--17] in order to obtain the analytic
solutions.

We have succeeded in generating the Natanzon hypergeometric potentials \-by
making use of conformal mappings within the framework of position--dependent
mass (PDM) [18]. In this paper, we will explain the form in which the $%
\mathfrak{so(}2,1\mathfrak{)}$ Lie algebra can be enlarged as wide as
possible in order to generate the confluent Natanzon hypergeometric
potentials endowed with PDM. This is accomplished by scaling the resolvent
operator (Green's function) [19], in such a way that can be related to
nonrelativistic PDM Schr\"{o}dinger equation [20--24]. This explicit
connection between the resolvent operator transformations to PDM quantum
systems.

The organization of the present paper is as follows. We introduce in section
2 a specific differential realization of the $\mathfrak{so}\left( 2,1\right) 
$ Lie algebra. Then applying a scaling in the resolvent operator, through a
scale function that preserves the wavefunctions, will map it into the wave
equation endowed with position--dependent mass. This is possible if the
scale function is identified to the mass distribution. This correspondence
gives, in section 3, the (effective) potential expression for confluent
Natanzon potentials in the framework of position--dependent mass. The last
section is kept for our concluding remarks and an appendix was added in
order to\ point out the correct use of the scalar product under this scaling.

\section{Resolvent operator transformations and $\mathfrak{so}\left(
2,1\right) $ algebra}

We are going to build a realization of the $\mathfrak{so}\left( 2,1\right) $
Lie algebra according to [12]. Let us start from a couple of operators $a$
and $b$ satisfying the commutation relation%
\begin{equation}
\left[ a,b\right] =1,  \tag{1}
\end{equation}%
and defining the following realization for the generators $J_{k}\left(
k=0,1,2\right) $%
\begin{eqnarray}
J_{0} &=&\frac{1}{4}\left[ -a^{2}+\left( 4c+\frac{3}{4}\right) b^{-2}+b^{2}%
\right] ,  \TCItag{2.1} \\
J_{1} &=&\frac{1}{4}\left[ -a^{2}+\left( 4c+\frac{3}{4}\right) b^{-2}-b^{2}%
\right] ,  \TCItag{2.2} \\
J_{2} &=&-\frac{i}{4}\left( 2ba+1\right) ,  \TCItag{2.3}
\end{eqnarray}%
where $c$ are eigenvalues of the Casimir's operator of the algebra. The
relations (2) satisfy the commutation relations%
\begin{equation}
\left[ J_{0},J_{1}\right] =iJ_{2},\quad \left[ J_{2},J_{0}\right]
=iJ_{1},\quad \left[ J_{1},J_{2}\right] =-iJ_{0}.  \tag{3}
\end{equation}

It might be of interest to consider the operators $a$ and $b$ in terms of a
variable $x$ as $a\equiv \frac{d}{dx}+K\left( x\right) $ and $b\equiv x$.
Performing the point canonical transformation $x=\sqrt{\xi \left( u\right) }$%
, the operators $a$ and $b$ become%
\begin{equation}
a=\frac{2\sqrt{\xi \left( u\right) }}{\xi ^{\prime }\left( u\right) }\frac{d%
}{du}-\frac{1}{2}\left( \frac{2\sqrt{\xi \left( u\right) }}{\xi ^{\prime
}\left( u\right) }\right) ^{\prime },\qquad b=\sqrt{\xi \left( u\right) }. 
\tag{4}
\end{equation}%
where here and below the prime stands for derivatives in variable $u$. In
terms of this realization, the generators (2) become [12]%
\begin{eqnarray}
J_{0} &=&-\frac{\xi \left( u\right) }{\xi ^{\prime 2}\left( u\right) }\frac{%
d^{2}}{du^{2}}-\frac{1}{2}\frac{\xi ^{\prime \prime \prime }\left( u\right)
\xi \left( u\right) }{\xi ^{\prime 3}\left( u\right) }+\frac{3}{4}\frac{\xi
^{\prime \prime 2}\left( u\right) \xi \left( u\right) }{\xi ^{\prime
4}\left( u\right) }+\frac{c}{\xi \left( u\right) }+\frac{\xi \left( u\right) 
}{4},  \TCItag{5.1} \\
J_{1} &=&-\frac{\xi \left( u\right) }{\xi ^{\prime 2}\left( u\right) }\frac{%
d^{2}}{du^{2}}-\frac{1}{2}\frac{\xi ^{\prime \prime \prime }\left( u\right)
\xi \left( u\right) }{\xi ^{\prime 3}\left( u\right) }+\frac{3}{4}\frac{\xi
^{\prime \prime 2}\left( u\right) \xi \left( u\right) }{\xi ^{\prime
4}\left( u\right) }+\frac{c}{\xi \left( u\right) }-\frac{\xi \left( u\right) 
}{4},  \TCItag{5.2} \\
J_{2} &=&-i\frac{\xi \left( u\right) }{\xi ^{\prime }\left( u\right) }\frac{d%
}{du}-\frac{i}{2}\frac{\xi ^{\prime \prime }\left( u\right) \xi \left(
u\right) }{\xi ^{\prime 2}\left( u\right) }.  \TCItag{5.3}
\end{eqnarray}

The Green's function for a wide class of potentials is given following
expression%
\begin{equation}
\left( H-E\right) G\left( u,\overline{u}\right) =\delta \left( u-\overline{u}%
\right) ,  \tag{6}
\end{equation}%
and due to the form of the generators in (5), it is easy to give the
resolvent operator [19]%
\begin{eqnarray}
\Omega \left( E\right) &=&\frac{\xi \left( u\right) }{\xi ^{\prime 2}\left(
u\right) }\left( H-E\right)  \notag \\
&=&q_{0}+\dsum\limits_{k=0}^{2}p_{k}J_{k},  \TCItag{7}
\end{eqnarray}%
where $q_{0}$ and $p_{k}\left( k=0,1,2\right) $ are arbitrary parameters
that will be fixed below.

The realization in (7) may be implemented by scaling the resolvent operator
on the generators (2) in such a way to explicit connection between the
resolvent operator (7), and then the spectrum generating algebra method, to
PDM Schr\"{o}dinger equation. Under such transformations, the operator $%
\Omega \left( E\right) $ becomes%
\begin{eqnarray}
\overline{\Omega }\left( E\right) &=&\left(
q_{0}+\dsum\limits_{k=0}^{2}p_{k}J_{k}\right) P\left( u,\overline{u}\right) 
\notag \\
&=&P\left( u,\overline{u}\right) \left(
q_{0}+\dsum\limits_{k=0}^{2}p_{k}T_{k}\right) ,  \notag \\
&=&P\left( u,\overline{u}\right) \frac{\xi \left( u\right) }{\xi ^{\prime
2}\left( u\right) }\left( \widetilde{H}-E\right)  \TCItag{8}
\end{eqnarray}%
where $P\left( u,\overline{u}\right) =P_{1}\left( u\right) P_{2}\left( 
\overline{u}\right) $ is called \textit{two--point scale }function, and $%
T_{k}$ are generators which are determined through identification.

Under this change and by identifying both sides of (8) in terms of
parameters $q_{0}$ and $p_{k}$, the generators $T_{k}$ read as%
\begin{eqnarray}
T_{0} &=&-\frac{\xi }{\xi ^{\prime 2}}\frac{d^{2}}{du^{2}}-2\frac{\xi }{\xi
^{\prime 2}}\frac{P^{\prime }}{P}\frac{d}{du}-\frac{\xi }{\xi ^{\prime 2}}%
\frac{P^{\prime \prime }}{P}-\frac{1}{2}\frac{\xi ^{\prime \prime \prime
}\xi }{\xi ^{\prime 3}}+\frac{3}{4}\frac{\xi ^{\prime \prime 2}\xi }{\xi
^{\prime 4}}+\frac{c}{\xi }+\frac{\xi }{4},  \TCItag{9.1} \\
T_{1} &=&-\frac{\xi }{\xi ^{\prime 2}}\frac{d^{2}}{du^{2}}-2\frac{\xi }{\xi
^{\prime 2}}\frac{P^{\prime }}{P}\frac{d}{du}-\frac{\xi }{\xi ^{\prime 2}}%
\frac{P^{\prime \prime }}{P}-\frac{1}{2}\frac{\xi ^{\prime \prime \prime
}\xi }{\xi ^{\prime 3}}+\frac{3}{4}\frac{\xi ^{\prime \prime 2}\xi }{\xi
^{\prime 4}}+\frac{c}{\xi }-\frac{\xi }{4},  \TCItag{9.2} \\
T_{2} &=&-i\frac{\xi }{\xi ^{\prime }}\frac{d}{du}-i\frac{\xi }{\xi ^{\prime
}}\frac{P^{\prime }}{P}-\frac{i}{2}\frac{\xi ^{\prime \prime }\xi }{\xi
^{\prime 2}},  \TCItag{9.3}
\end{eqnarray}%
where $P$ here is referred to the function $P_{1}\left( u\right) $ due to
differentiation with respect to the variable $u$.

By introducing the eigenfunctions%
\begin{equation}
\psi \left( u\right) =2\frac{\xi ^{2}\left( u\right) }{\xi ^{\prime 2}\left(
u\right) }\phi \left( u\right) ,  \tag{10}
\end{equation}%
the operator $\overline{\Omega }\left( E\right) $ becomes after some
straightforward algebras%
\begin{eqnarray}
\overline{\Omega }\left( E\right) &=&P\left( u,\overline{u}\right) \frac{\xi 
}{\xi ^{\prime 2}}\left[ -2p_{+}\frac{\xi ^{2}}{\xi ^{\prime 2}}\frac{d^{2}}{%
du^{2}}-\frac{\xi }{\xi ^{\prime }}\left\{ 4p_{+}\left( 2+\frac{P^{\prime }}{%
P}\frac{\xi }{\xi ^{\prime }}-2\frac{\xi ^{\prime \prime }\xi }{\xi ^{\prime
2}}\right) +2ip_{2}\xi \right\} \frac{d}{du}\right.  \notag \\
&&-p_{+}\left( 8\frac{P^{\prime }}{P}\frac{\xi }{\xi ^{\prime }}+4-2c+2\frac{%
P^{\prime \prime }}{P}\frac{\xi ^{2}}{\xi ^{\prime 2}}-8\frac{P^{\prime }}{P}%
\frac{\xi ^{2}\xi ^{\prime \prime }}{\xi ^{\prime 3}}-12\frac{\xi \xi
^{\prime \prime }}{\xi ^{\prime 2}}+\frac{21}{2}\frac{\xi ^{2}\xi ^{\prime
\prime 2}}{\xi ^{\prime 4}}-3\frac{\xi ^{2}\xi ^{\prime \prime \prime }}{\xi
^{\prime 3}}\right)  \notag \\
&&\left. +2q_{0}\xi +\frac{p_{-}}{2}\xi ^{2}-2ip_{2}\left( \frac{P^{\prime }%
}{P}\frac{\xi ^{2}}{\xi ^{\prime }}+2\xi -\frac{3}{2}\frac{\xi ^{\prime
\prime }\xi ^{2}}{\xi ^{\prime 2}}\right) \right] ,  \TCItag{11}
\end{eqnarray}%
where $p_{\pm }=p_{0}\pm p_{1}$.

Without any loss of generality, let us take the appropriate--parameter
choices on $p_{\pm }$ and $p_{2}$ such as%
\begin{equation}
p_{+}\equiv p_{0}+p_{1}=\frac{\alpha }{2},\quad p_{-}\equiv p_{0}-p_{1}=%
\frac{\beta }{2},\quad p_{2}\equiv 0,  \tag{12}
\end{equation}%
then the operator $\overline{\Omega }\left( E\right) $ in (11) becomes%
\begin{eqnarray}
\overline{\Omega }\left( E\right) &\equiv &P\left( u,\overline{u}\right) 
\frac{\xi }{\xi ^{\prime 2}}\left( \widetilde{H}-E\right)  \notag \\
&=&P\left( u,\overline{u}\right) \frac{\xi }{\xi ^{\prime 2}}\left[ -\alpha 
\frac{\xi ^{2}}{\xi ^{\prime 2}}\frac{d^{2}}{du^{2}}-\alpha \frac{\xi }{\xi
^{\prime }}\left( 4+2\frac{P^{\prime }}{P}\frac{\xi }{\xi ^{\prime }}-4\frac{%
\xi ^{\prime \prime }\xi }{\xi ^{\prime 2}}\right) \frac{d}{du}\right. 
\notag \\
&&+2q_{0}\xi +\frac{\beta }{4}\xi ^{2}-4\alpha \frac{P^{\prime }}{P}\frac{%
\xi }{\xi ^{\prime }}-2\alpha +\alpha c-\alpha \frac{P^{\prime \prime }}{P}%
\frac{\xi ^{2}}{\xi ^{\prime 2}}+4\alpha \frac{P^{\prime }}{P}\frac{\xi
^{2}\xi ^{\prime \prime }}{\xi ^{\prime 3}}+6\alpha \frac{\xi \xi ^{\prime
\prime }}{\xi ^{\prime 2}}  \notag \\
&&\left. -\frac{21\alpha }{4}\frac{\xi ^{2}\xi ^{\prime \prime 2}}{\xi
^{\prime 4}}+\frac{3\alpha }{2}\frac{\xi ^{2}\xi ^{\prime \prime \prime }}{%
\xi ^{\prime 3}}\right] .  \TCItag{13}
\end{eqnarray}

On the other hand, in the case of varying mass which will be denoted by $%
M\left( u\right) =m_{0}m\left( u\right) $, the Hamiltonian introduced by von
Roos [20] reads as%
\begin{equation}
H_{vR}=\frac{1}{4}\left( m^{\eta }\boldsymbol{p}m^{\epsilon }\boldsymbol{p}%
m^{\rho }+m^{\rho }\boldsymbol{p}m^{\epsilon }\boldsymbol{p}m^{\eta }\right)
+V_{\text{eff}}\left( u\right) ,  \tag{14}
\end{equation}%
where $\eta $, $\epsilon $, and $\rho $ are three parameters which obey to
the restriction $\eta +\epsilon +\rho =-1$ in order to grant the classical
limit. Here $\boldsymbol{p}\left( =-i\frac{d}{du}\right) $ is a momentum
with $\hbar =m_{0}=1$, and $m\left( u\right) $ is dimensionless--real valued
mass.

Applying the eigenfunctions [18,23]%
\begin{eqnarray}
\Psi \left( u\right) &\equiv &m\left( u\right) \psi \left( u\right)  \notag
\\
&=&2m\left( u\right) \frac{\xi ^{2}\left( u\right) }{\xi ^{\prime 2}\left(
u\right) }\phi \left( u\right)  \TCItag{15}
\end{eqnarray}%
the resolvent operator of $H_{vR}$ reads [18]%
\begin{eqnarray}
H_{vR}-E &=&-\frac{\xi ^{2}}{\xi ^{\prime 2}}\frac{d^{2}}{du^{2}}-\frac{\xi 
}{\xi ^{\prime }}\left( 4+\frac{m^{\prime }}{m}\frac{\xi }{\xi ^{\prime }}-%
\frac{4\xi \xi ^{\prime \prime }}{\xi ^{\prime 2}}\right) \frac{d}{du}+\frac{%
2\xi }{\xi ^{\prime 2}}\left( 3\xi ^{\prime \prime }+\frac{\xi \xi ^{\prime
\prime \prime }}{\xi ^{\prime }}-\frac{3\xi \xi ^{\prime \prime 2}}{\xi
^{\prime 2}}\right)  \notag \\
&&+\frac{m^{\prime }}{m}\frac{\xi ^{2}}{\xi ^{\prime 2}}\left( 2\frac{\xi
\xi ^{\prime \prime }-\xi ^{\prime 2}}{\xi \xi ^{\prime }}+\frac{\epsilon -1%
}{2}\frac{m^{\prime \prime }}{m^{\prime }}-\left( 1+\eta \right) \left( \eta
+\epsilon \right) \frac{m^{\prime }}{m}\right) -2  \notag \\
&&+2m\frac{\xi ^{2}}{\xi ^{\prime 2}}\left( V_{\text{eff}}\left( x\right)
-E\right) .  \TCItag{16}
\end{eqnarray}

Then comparing (13) to (16) in their differential terms, we obtain%
\begin{equation}
P\left( u,\overline{u}\right) =\sqrt{m\left( u\right) m\left( \overline{u}%
\right) },  \tag{17}
\end{equation}%
leading to defining the one--point scale function $P\left( u,u\right) \equiv
P\left( u\right) =m\left( u\right) $ as $u=\overline{u}$. By substituting
(17) into the remainder term leads, after long calculations, to the
following general expression for the potential%
\begin{equation}
V\left[ \xi \left( u\right) \right] -E=\frac{1}{2m\left( u\right) }\frac{\xi
^{\prime 2}\left( u\right) }{\xi ^{2}\left( u\right) }\left( 2q_{0}\xi
\left( u\right) +\frac{\beta }{4}\xi ^{2}\left( u\right) +c\right) -\frac{1}{%
4m\left( u\right) }\left\{ \xi \left( u\right) ,u\right\} _{\text{S}}, 
\tag{18}
\end{equation}%
and which corresponds to the specific--parameter choice $\alpha =1$. Here, $%
\left\{ \xi \left( u\right) ,u\right\} _{\text{S}}$ is the Schwarz
derivative of $\xi \left( u\right) $ with respect to $u$ defined by%
\begin{equation}
\left\{ \xi \left( u\right) ,u\right\} _{\text{S}}=\frac{\xi ^{\prime \prime
}\left( u\right) }{\xi ^{\prime }\left( u\right) }\left( \frac{\xi ^{\prime
\prime \prime }\left( u\right) }{\xi ^{\prime \prime }\left( u\right) }-%
\frac{3}{2}\frac{\xi ^{\prime \prime }\left( u\right) }{\xi ^{\prime }\left(
u\right) }\right) ,  \tag{19}
\end{equation}%
and $V\left[ \xi \left( u\right) \right] =V_{\text{eff}}\left[ \xi \left(
u\right) \right] -\mathcal{V}_{\text{m}}^{\left( \eta ,\epsilon \right)
}\left( u\right) $, with%
\begin{equation}
\mathcal{V}_{\text{m}}^{\left( \eta ,\epsilon \right) }\left( u\right) =%
\frac{\left( 1+2\eta \right) ^{2}+4\epsilon \left( 1+\eta \right) }{8}\frac{%
m^{\prime 2}\left( u\right) }{m^{3}\left( u\right) }-\frac{\epsilon }{4}%
\frac{m^{\prime \prime }\left( u\right) }{m^{2}\left( u\right) },  \tag{20}
\end{equation}%
which is attributed to the dependence of the mass $m$ on$\ u$.

\section{Generating confluent Natanzon potentials}

Since we have to get a constant term in left--hand side of (18), we must
divide the right--hand side into two parts: the former represents a constant
term and which is identified to the energy $E$, while the latter corresponds
to the potential $V\left[ \xi \left( u\right) \right] $. This condition
amounts to setting up differential equation for $\xi \left( u\right) $ and
satisfying%
\begin{equation}
\frac{\xi ^{\prime 2}\left( u\right) }{2m\left( u\right) }\frac{\lambda
_{2}\xi ^{2}\left( u\right) +\lambda _{1}\xi \left( u\right) +\lambda _{0}}{%
4\xi ^{2}\left( u\right) }=1,  \tag{21}
\end{equation}%
where $\lambda _{i}\left( i=0,1,2\right) $ are arbitrary parameters.
Inserting (21) into (18), we get%
\begin{equation}
V\left[ \xi \left( u\right) \right] -E=\frac{\beta \xi ^{2}\left( u\right)
+8q_{0}\xi \left( u\right) +4c}{\lambda _{2}\xi ^{2}\left( u\right) +\lambda
_{1}\xi \left( u\right) +\lambda _{0}}-\frac{1}{4m\left( u\right) }\left\{
\xi \left( u\right) ,u\right\} _{\text{S}}.  \tag{22}
\end{equation}

However in order to get a constant term in right--hand side of (22), it is
useful to make some particularizations on the parameters $\beta $, $q_{0}$,
and $c$. As a consequence, it will turn out to be useful to express the
resolvent operator $\left( \widetilde{H}-E\right) $ in (13) as the
exponential function of the generator $T_{2}$, since $T_{2}$ is essentially
a generator of \textit{scale} transformations [25].

To this end, taking into account (12), the resolvent operator (8) reads%
\begin{equation}
\left( \widetilde{H}-E\right) \psi \left( u\right) =\frac{\xi ^{\prime
2}\left( u\right) }{\xi \left( u\right) }\left[ 2\left( \beta +1\right)
T_{0}-2\left( \beta -1\right) T_{1}+8q_{0}\right] \psi \left( u\right) =0. 
\tag{23}
\end{equation}

It follows from the identity [26]%
\begin{eqnarray}
\func{e}^{i\theta T_{2}}T_{0}\func{e}^{-i\theta T_{2}} &\equiv
&\dsum\limits_{k=0}^{\infty }\frac{1}{n!}\left[ i\theta T_{2},T_{0}\right]
_{\left( n\right) }  \notag \\
&=&T_{0}\cosh \theta -T_{1}\sinh \theta ,  \TCItag{24}
\end{eqnarray}%
where%
\begin{eqnarray}
\left[ i\theta T_{2},T_{0}\right] _{\left( n\right) } &=&\left[ i\theta
T_{2},\left[ i\theta T_{2},T_{0}\right] _{\left( n-1\right) }\right]  \notag
\\
&=&\left[ i\theta T_{2},\left[ i\theta T_{2},\left[ i\theta T_{2},T_{0}%
\right] _{\left( n-2\right) }\right] \right]  \notag \\
&=&\cdots  \TCItag{25}
\end{eqnarray}%
that the right--hand side of (23) becomes%
\begin{equation}
\mathcal{T}_{0}\psi \left( u\right) \equiv \func{e}^{i\delta T_{2}}T_{0}%
\func{e}^{-i\delta T_{2}}\psi \left( u\right) =-\frac{2q_{0}}{\sqrt{\beta }}%
\psi \left( u\right) ,  \tag{26}
\end{equation}%
where $\theta $ and $\delta $ are expressed in terms of $\beta $ as%
\begin{equation}
\text{ }\theta =\frac{\beta -1}{\beta +1},\quad \text{and\quad }\delta =%
\frac{1}{2}\ln \beta ,  \tag{27}
\end{equation}%
and are allowed to depend on $n$ and $c$.

Since $J_{0}$ is the only compact generator with eigenvalues $\left\langle
J_{0}\right\rangle =j_{0}$, then the allowed values of $j_{0}$ are related
to $n$ and $c$ through [12]%
\begin{equation}
j_{0}=n+\frac{1}{2}\left( 1+\sqrt{1+4c}\right) ,  \tag{28}
\end{equation}%
where $n=0,1,2,\ldots $ Combining (28) to (26), we get the basic equation%
\begin{equation}
q_{0}+\frac{\sqrt{\beta }}{4}\left( 2n+1+\sqrt{1+4c}\right) =0.  \tag{29}
\end{equation}

It is then obvious that the simplest set of parameters associated with $%
\beta $, $q_{0}$ and $c$ and compatible with (29) is%
\begin{equation}
\beta =a^{2},\quad q_{0}=-\frac{a}{2}\left( \frac{b}{2}+n\right) ,\quad c=%
\frac{b}{2}\left( \frac{b}{2}-1\right) ,  \tag{30}
\end{equation}%
where the equation (29) can be obtained using the equalities (30) by the
straight substitution.

Equation (22) becomes%
\begin{equation}
V\left[ \xi \left( u\right) \right] -E=\frac{a^{2}\xi ^{2}\left( u\right)
-2a\left( 2n+b\right) \xi \left( u\right) +b\left( b-2\right) }{\lambda
_{2}\xi ^{2}\left( u\right) +\lambda _{1}\xi \left( u\right) +\lambda _{0}}-%
\frac{1}{4m\left( u\right) }\left\{ \xi \left( u\right) ,u\right\} _{\text{S}%
}.  \tag{31}
\end{equation}%
which in particular case requires that coefficients in numerator are linear
with those of denominator with respect to $E$ following%
\begin{eqnarray}
a^{2} &=&-\lambda _{2}E+\sigma _{\beta },  \TCItag{32.1} \\
2a\left( 2n+b\right) &=&\lambda _{1}E-\sigma _{q_{0}},  \TCItag{32.2} \\
b\left( b-2\right) &=&-\lambda _{0}E+\sigma _{c},  \TCItag{32.3}
\end{eqnarray}%
leading to express the potential $V\left( u\right) $ into%
\begin{equation}
V\left[ \xi \left( u\right) \right] =\frac{\sigma _{\beta }\xi ^{2}\left(
u\right) +\sigma _{q_{0}}\xi \left( u\right) +\sigma _{c}}{R\left[ \xi
\left( u\right) \right] }-\frac{1}{4m\left( u\right) }\left\{ \xi \left(
u\right) ,u\right\} _{\text{S}},  \tag{33}
\end{equation}%
with $R\left( \xi \right) =\lambda _{2}\xi ^{2}+\lambda _{1}\xi +\lambda
_{0} $. For convenience, we introduce the auxiliary function from (21) as
[12,18]%
\begin{equation}
\mathfrak{S}\left[ \xi \left( u\right) \right] \equiv \frac{\xi ^{\prime
2}\left( u\right) }{2m\left( u\right) }=\frac{4\xi ^{2}\left( u\right) }{R%
\left[ \xi \left( u\right) \right] },  \tag{34}
\end{equation}%
therefore the Schwarz derivative--term can be expressed in terms of $\xi
\left( u\right) $%
\begin{equation}
\frac{1}{4m\left( u\right) }\left\{ \xi \left( u\right) ,u\right\} _{\text{S}%
}=-\frac{1}{R}-\frac{\xi ^{2}\overset{..}{R}+\xi \overset{.}{R}}{R^{2}}+%
\frac{5}{4}\xi ^{2}\frac{\overset{.}{R}^{2}}{R^{3}}-U_{\text{m}}\left(
u\right) ,  \tag{35}
\end{equation}%
where the dot refers to the derivative with respect to $\xi $ and $U_{\text{m%
}}\left( u\right) =\frac{5}{32}\frac{m^{\prime 2}\left( u\right) }{%
m^{3}\left( u\right) }-\frac{1}{8}\frac{m^{\prime \prime }\left( u\right) }{%
m^{2}\left( u\right) }$.

Inserting (35) into (33), taking into account the first and the second
derivative, we get%
\begin{equation}
V\left( u\right) =\frac{\sigma _{\beta }\xi ^{2}+\sigma _{q_{0}}\xi +\sigma
_{c}+1}{R\left( \xi \right) }+\frac{4\lambda _{2}\xi ^{2}+\lambda _{1}\xi }{%
R^{2}\left( \xi \right) }-\frac{5}{4}\xi ^{2}\frac{\left( 4\lambda _{2}\xi
+\lambda _{1}\right) ^{2}}{R^{3}\left( \xi \right) },  \tag{36}
\end{equation}%
where a effective--mass potential is now given 
\begin{eqnarray}
\mathcal{U}_{\text{m}}^{\left( \eta ,\epsilon \right) }\left( u\right) &=&%
\mathcal{V}_{\text{m}}^{\left( \eta ,\epsilon \right) }\left( u\right) +U_{%
\text{m}}\left( u\right)  \notag \\
&=&\frac{4\left( 1+2\eta \right) ^{2}+16\epsilon \left( 1+\eta \right) +5}{32%
}\frac{m^{\prime 2}\left( u\right) }{m^{3}\left( u\right) }-\frac{2\epsilon
+1}{8}\frac{m^{\prime \prime }\left( u\right) }{m^{2}\left( u\right) }. 
\TCItag{37}
\end{eqnarray}

Now consider simultaneous addition and subtraction of $\lambda
_{1}^{2}-\lambda _{0}\lambda _{2}$, this procedure allows us to get $R\left(
\xi \right) $ in the numerator of the third term. We obtain after some
straightforward algebras%
\begin{equation}
V\left( u\right) =\frac{\sigma _{\beta }\xi ^{2}+\sigma _{q_{0}}\xi +\sigma
_{c}+1}{R\left( \xi \right) }+\frac{\lambda _{1}\xi -\lambda _{2}\xi ^{2}}{%
R^{2}\left( \xi \right) }-\frac{5}{4}\xi ^{2}\frac{\Delta }{R^{3}\left( \xi
\right) },  \tag{38}
\end{equation}%
where $\Delta =\lambda _{1}^{2}-4\lambda _{0}\lambda _{2}$. We recognize in
(38) the general expression of the confluent Natanzon potentials. Note that
the potential deduced is completely independent of ambiguity parameters $%
\eta $, $\epsilon $, and $\rho $. However, the potential differs
substantially from that usually known in the literature [11,12].

\subsection{Energy eigenvalues $E_{n}$}

The corresponding energy eigenvalues of the confluent Natanzon potentials
can be determined from (32). In fact, proceeding to squaring (32.a) and
(32.c) and by inserting it into (32.b), we see then that%
\begin{equation}
2n+1=\frac{-\sigma _{q_{0}}+\lambda _{1}E_{n}}{2\sqrt{\sigma _{\beta
}-\lambda _{2}E_{n}}}-\sqrt{1+\sigma _{c}-\lambda _{0}E_{n}},  \tag{39}
\end{equation}%
is an equality of the fourth degree in $E_{n}$ and agree with that obtained
by [11,12]. It is obvious that parameters $a$ and $b$ depend on quantum
number $n$.

\subsection{Wavefunctions $\overline{\protect\psi }_{n}\left[ \protect\xi %
\left( u\right) \right] $}

The two sets of operators $T_{i}$ and $\mathcal{T}_{i}$ are obviously
different but have the same effects on the wavefunctions. Using (24) and
(26), it turns out that the parameter $\sqrt{\beta }=a_{n}$ plays the role
of scaling parameter in such a way that $\xi \left( u\right) $ can be mapped
into $a_{n}\xi \left( u\right) $, but does not affect the general form of
wavefunctions.

The wavefunctions belong to the Hilbert space, not with respect to the usual
scalar product of quantum mechanics but with respect to the new scalar
product (see the appendix)%
\begin{eqnarray}
\left\langle \overline{\psi }_{j}|\overline{\psi }_{i}\right\rangle _{W}
&\equiv &\dint\nolimits_{-\infty }^{+\infty }\overline{\psi }_{i}\left(
u\right) \overline{\psi }_{j}^{\ast }\left( u\right) du  \notag \\
&=&\dint\nolimits_{-\infty }^{+\infty }\chi _{i}\left( u\right) \frac{1}{%
W\left( u\right) }\chi _{j}^{\ast }\left( u\right) du,  \TCItag{40}
\end{eqnarray}%
where $\overline{\psi }_{i}\left( u\right) =\frac{1}{\sqrt{W\left( u\right) }%
}\chi _{i}\left( u\right) \left( i=1,2,\ldots \right) $ and $W\left(
u\right) \equiv \frac{1}{P\left( u\right) }=\frac{1}{m\left( u\right) }$
represents the \textit{weight factor}.

Then, and following [9], the assumption of an unnormalized wavefunction $%
\overline{\psi }_{n}\left[ \xi \left( u\right) \right] $ gets transformed as%
\begin{eqnarray}
\overline{\psi }_{n}\left[ \xi \left( u\right) \right] &\sim &\sqrt{\frac{%
m\left( u\right) }{\xi ^{\prime }\left( u\right) }}\xi ^{b_{n}/2}\left(
u\right) \func{e}^{-\xi \left( u\right) /2}{}_{1}F_{1}\left(
-n;b_{n};a_{n}\xi \left( u\right) \right)  \notag \\
&=&m^{1/4}\left( u\right) R^{1/4}\left( u\right) \xi ^{\left( b_{n}-1\right)
/2}\left( u\right) \func{e}^{-\xi \left( u\right) /2}{}_{1}F_{1}\left(
-n;b_{n};a_{n}\xi \left( u\right) \right) ,  \TCItag{41}
\end{eqnarray}%
where the identity (34) is used. The wavefunction (41) has an extra term $%
\sqrt{m\left( u\right) }$ compared with that for the constant mass (see the
appendix). This achieves the bound--state problem for the confluent Natanzon
potentials.

\section{Conclusion}

We have investigated another method to generate the confluent Natanzon
potentials and its bound--states. The method consists to enlarge the $%
\mathfrak{so}\left( 2,1\right) $ Lie group generators in order to relate a
resolvent operator transformations to the position--dependent mass Schr\"{o}%
dinger equation. We have shown that if a particular correspondence between
the scale function and mass distribution exists, i.e. $P\left( u,\overline{u}%
\right) =\sqrt{m\left( u\right) m\left( \overline{u}\right) }$, then there
is a connection procedure between the resolvent operator transformations and
position--dependent mass Schr\"{o}dinger equation leading to generate the
confluent Natanzon potentials and their bound--states.\- We have also
pointed out the correct use of the $W\left( u\right) $--scalar product when
dealing with position--dependent mass.

\section{Appendix: The correct scalar product}

As outlined in the introduction, this appendix was added in order to bring
some mathematical details were not approached in subsection 3.2 relating to
the correct use of the scalar product under the $P\left( u\right) $--scaling.

The time--dependent Schr\"{o}dinger equation for the functions $\psi \left(
u,t\right) $ and $\psi ^{\ast }\left( u,t\right) $ and associated to the
Hamiltonian (14) read, respectively, as [27]%
\begin{eqnarray}
i\frac{\partial \psi \left( u,t\right) }{\partial t} &=&-\frac{\partial }{%
\partial u}\frac{1}{2m\left( u\right) }\frac{\partial }{\partial u}\psi
\left( u,t\right) +V_{\text{eff}}\left( u\right) \psi \left( u,t\right) , 
\TCItag{A.1} \\
-i\frac{\partial \psi ^{\ast }\left( u,t\right) }{\partial t} &=&-\frac{%
\partial }{\partial u}\frac{1}{2m\left( u\right) }\frac{\partial }{\partial u%
}\psi ^{\ast }\left( u,t\right) +V_{\text{eff}}\left( u\right) \psi ^{\ast
}\left( u,t\right) ,  \TCItag{A.2}
\end{eqnarray}%
where $V_{\text{eff}}\left( u\right) \in 
\mathbb{R}
$.

Let $\psi _{1}\left( u,t\right) $ and $\psi _{2}^{\ast }\left( u,t\right) $
be the solutions of (A.1) and (A.2) respectively. By multiplying (A.1) by $%
W\left( u\right) \psi _{2}^{\ast }\left( u,t\right) $ and (A.2) by $W\left(
u\right) \psi _{1}\left( u,t\right) $ on the left--hand side, we end up on
subtracting both equations to the equality%
\begin{multline}
i\frac{\partial }{\partial t}\left[ \psi _{1}\left( u,t\right) W\left(
u\right) \psi _{2}^{\ast }\left( u,t\right) \right] =  \notag \\
-W\left( u\right) \left[ \psi _{2}^{\ast }\left( u,t\right) \frac{\partial }{%
\partial u}\frac{1}{2m\left( u\right) }\frac{\partial }{\partial u}\psi
_{1}\left( u,t\right) -\psi _{1}\left( u,t\right) \frac{\partial }{\partial u%
}\frac{1}{2m\left( u\right) }\frac{\partial }{\partial u}\psi _{2}^{\ast
}\left( u,t\right) \right] ,  \tag{A.3}
\end{multline}%
due to the fact that $W\left( u\right) $ is independent of time $t$. Then
(A.3) can be recast into the differential form as%
\begin{multline}
i\frac{\partial }{\partial t}\left[ \psi _{1}\left( u,t\right) W\left(
u\right) \psi _{2}^{\ast }\left( u,t\right) \right]  \notag \\
=-\frac{N}{2}W\left( u\right) \frac{\partial }{\partial u}\left\{ W\left(
u\right) \left[ \psi _{2}^{\ast }\left( u,t\right) \frac{\partial \psi
_{1}\left( u,t\right) }{\partial u}-\psi _{1}\left( u,t\right) \frac{%
\partial \psi _{2}^{\ast }\left( u,t\right) }{\partial u}\right] \right\} , 
\tag{A.4}
\end{multline}%
with a specific choice of the weight function $W\left( u\right) $ chosen as%
\begin{equation}
W\left( u\right) \equiv \frac{1}{N}\frac{1}{P\left( u\right) }=\frac{1}{N}%
\frac{1}{m\left( u\right) },\qquad \text{for any }N\in 
\mathbb{N}
-\left\{ 0\right\} .  \tag{A.5}
\end{equation}

Let us introduce the wavefunctions $\chi _{i}\left( u,t\right) \left(
i=1,2,\ldots \right) $ related to the eigenfunctions $\psi _{i}\left(
u,t\right) $ by%
\begin{equation}
\chi _{i}\left( u,t\right) =\sqrt{W\left( u\right) }\psi _{i}\left(
u,t\right) ,  \tag{A.6}
\end{equation}%
then (A.4) can be rewritten as%
\begin{equation}
i\frac{\partial }{\partial t}\left( \chi _{1}\left( u,t\right) \frac{1}{%
W\left( u\right) }\chi _{2}^{\ast }\left( u,t\right) \right) =-\frac{N}{2}%
\frac{\partial }{\partial u}\left( \chi _{2}^{\ast }\left( u,t\right) \frac{%
\partial \chi _{1}\left( u,t\right) }{\partial u}-\chi _{1}\left( u,t\right) 
\frac{\partial \chi _{2}^{\ast }\left( u,t\right) }{\partial u}\right) . 
\tag{A.7}
\end{equation}

By integrating (A.7) with respect to the variable $u$, we get%
\begin{equation}
\frac{\partial }{\partial t}\int_{-\infty }^{+\infty }\chi _{1}\left(
u,t\right) \frac{1}{W\left( u\right) }\chi _{2}^{\ast }\left( u,t\right)
du=0,  \tag{A.8}
\end{equation}%
where, as is excepted for bound--state wavefunctions, $\chi _{1}\left(
u,t\right) \rightarrow 0$ and $\chi _{2}^{\ast }\left( u,t\right)
\rightarrow 0$ as $u\rightarrow \pm \infty $. It becomes clear that these
wavefunctions are not orthonormal as they stand; rather they are orthonormal
with respect to \textit{the weight factor} $\frac{1}{W\left( u\right) }$
defined in (A.5). Dealing only with spatial wavefunctions, i.e. $\chi
_{i}\left( u\right) $, (A.8) leads to the new scalar product%
\begin{equation}
\left\langle \overline{\psi }_{2}|\overline{\psi }_{1}\right\rangle
_{W}\equiv \int_{-\infty }^{+\infty }\chi _{1}\left( u\right) \frac{1}{%
W\left( u\right) }\chi _{2}^{\ast }\left( u\right) du,  \tag{A.9}
\end{equation}%
which is more general than that given usually in quantum mechanics. Through
(A.9), taken into account (A.5) and $N=1$, the eigenfunctions $\psi
_{i}\left( u\right) $ are mapped into%
\begin{equation}
\overline{\psi }_{i}\left( u\right) \equiv \frac{1}{\sqrt{W\left( u\right) }}%
\chi _{i}\left( u\right) =\sqrt{m\left( u\right) }\chi _{i}\left( u\right) ,
\tag{A.10}
\end{equation}%
which explain the manifestation of the term $\sqrt{m\left( u\right) }$ in
(41). Then due to the redefinition of the scalar product, it is obvious that 
$\chi _{i}\left( u\right) $ are the \textit{physical} wavefunctions, while $%
\psi _{i}\left( u\right) $ are called the \textit{group} wavefunctions.

\end{document}